# Simulation Studies for Electromagnetic Design of INO ICAL Magnet and its Response to Muons

S. P. Behera, M. S. Bhatia, V. M. Datar and A.K. Mohanty

*Abstract*— The iron calorimeter (ICAL) detector at the India-based Neutrino Observatory (INO) will be used to measure neutrino mass hierarchy. The magnet in the ICAL detector will be used to distinguish the µ⁻ and µ⁺ events induced by $\nu_\mu$ and anti-$\nu_\mu$ respectively. Due to the importance of the magnet in ICAL, an electromagnetic simulation has been carried out to study the B-field distribution in iron using various designs. The simulation shows better uniformity in the portion of the iron layer between the coils, which is bounded by regions which have lesser field strength as we move to the periphery of the iron layer. The ICAL magnet was configured to have a tiling structure that gave the minimum reluctance path while keeping a reasonably uniform field pattern. This translates into less Ampere-turns needed for generation of the required magnetic field. At low Ampere-turns, a larger fractional area with |B| ≥ 1 Tesla (T) can be obtained by using a soft magnetic material. A study of the effect of the magnetic field on muon trajectories has been carried out using GEANT4. For muons up to 20 GeV, the energy resolution improves as the magnetic field increases from 1.1 T to 1.8 T. The charge identification efficiency for muons was found to be more than 90% except for large zenith angles.

## I. INTRODUCTION

NEUTRINO physics today is an active field. The establishment of neutrino (ν) flavor oscillations and measurement of the neutrino parameters have been carried out by several neutrino experiments using atmospheric [1], solar [2], reactor [3 - 6] and accelerator [7 - 9] neutrino sources. Neutrino flavor oscillations require neutrinos to be massive and mixing between the 3 flavor eigenstates. This phenomenon cannot be understood within the standard model (SM) of the particle physics and is a hint to the physics beyond the SM. The determination of the sign of squared mass difference $(\Delta m^2_{32})$[1] is important for constructing the mass spectrum of neutrinos (mass hierarchy). This sign can be extracted using the earth matter effect, which is different for neutrinos and antineutrinos, in atmospheric and accelerator-based neutrino beams. The sensitivity of a proposed magnetized iron calorimetric detector, MONOLITH [10], to the neutrinos mass hierarchy was explored by Tabarelli de Fatis [11]. The issue of hierarchy of neutrino masses in a magnetized ICAL detector has been discussed in Refs. [12-14] by making use of the ratio (and difference) of events with µ⁻ and µ⁺ in the final state from muon neutrino and antineutrino interactions. The ICAL detector at INO [15], like MONOLITH and MINOS [16], with its charge identification capability can be used for studying mass hierarchy in neutrinos. The ICAL detector at the INO, which will host the biggest electromagnet in the world, leverages its ability to distinguish between muons of either charge to determine precisely the neutrino oscillation parameters such as $\Delta m^2$ and the mass hierarchy using atmospheric neutrinos [17]. The matter effect on neutrinos plays an important role for studying the mass hierarchy. Apart from ICAL at INO, future proposed detectors elsewhere aim to observe the earth matter effects on atmospheric neutrinos including the megaton water Cherenkov detectors such as Hyper-Kamiokande (HK) [18], large liquid argon detectors [19, 20] and a gigaton-class ice detector such as the Precision IceCube Next Generation Upgrade (PINGU) [21, 22]. The mass hierarchy and CP violation problem will also be addressed by accelerator based neutrino experiments such as NOνA [23], LBNE [24], and T2K [9].

The probability for oscillation of neutrinos from one flavor to another depends on the source to detector distance and energies of neutrinos. To measure the oscillation parameter more precisely, it is important that the neutrino energy and its incoming direction (from which the source to detector distance will be estimated) be accurately measured for each event. In order to estimate the distance traversed by the neutrino, it is also necessary to establish the flight direction (up vs. down) of the muons produced by the neutrinos with high efficiency. Techniques such as measuring the track curvature due to the presence of the magnetic field and the measurement of timing in successive detector layers can be used to achieve this. The ICAL detector has the ability to measure both these quantities. The energy of the neutrino is the sum of the hadronic and muon energies for a charged current interaction. It is difficult to reconstruct the energy for individual hadrons. However, the hit multiplicity of charged particles distinct from the muon track can be used to estimate the total energy of hadrons in the event [25]. The hadron momentum can also be estimated from its track length. The muon energy can be measured either by means of curvature in a magnetic field (magnetized iron), or from the range of muons stopping in the detector.

The ICAL detector will be placed in a cavern in order to reduce the cosmic ray background with a minimum all round rock cover of 1 km. The magnetized ICAL detector can be configured in one of the many possible geometries. For example, it could be a toroid (with or without iron), a layered cylindrical shape oriented in suitable fashion with a central conductor or a layered rectangular shaped one with embedded coils. We decided on the shape and dimensions of the ICAL

---

S. P. Behera , V. M. Datar and A.K. Mohanty are with the Nuclear Physics Division, Bhabha Atomic Research Centre, Mumbai - 400085, India and also with the Homi Bhabha National Institute, Anushaktinagar, Mumbai 400094, India (e-mail: shiba@barc.gov.in).

M. S. Bhatia is with the Laser and Plasma Technology Division, Bhabha Atomic Research Centre, Mumbai – 400085, and also with the Homi Bhabha National Institute, Anushaktinagar, Mumbai 400094, India.

detector keeping in mind the cavern dimensions of 132 m × 26 m × 32 m. The shape of ICAL is rectangular. To build a single detector of 51 kton would be impractical and hence a modular approach has been adopted with each module having a mass of ~17 kton and size of 16 m × 16 m × 14.5 m. Such a detector with a target mass of about 100 kton will be ideal to address the physics goals. The weight of ICAL is chosen by considering an acceptable event rate given the small interaction cross-section of neutrinos ~$10^{-42}$ m$^2$. To begin with, we proposed a detector of about 51 kton, which may be enlarged to double that size in stages.

In this paper we describe the studies carried out to optimize the design of the ICAL detector and find the parameters that have a bearing on power efficiency and uniformity of field within the ICAL magnet. We present results of electromagnetic simulations of the ICAL magnet using the finite element method based 3D commercial software. The motivation was to find the optimal slot configuration (through which pass the copper coils energizing ICAL), tiling of plates and to study the magnetic field strength and its uniformity for the baseline design and the effect of various kinds of departure from this design. These studies should help in obtaining the desired uniformity of as high a magnetic field as possible for the lowest power dissipation in the coil. In addition, the simulation studies have been carried out to see the effect of the magnetic field strength on muons using the object oriented "detector description and simulation tool" GEANT4 [26].

## II. UNIQUE REQUIREMENTS OF ICAL MAGNET

Each new application of electromagnets generates a need for new geometries, design goals and challenges. The ICAL magnet, which aims to detect neutrinos in the energy range of 1-20 GeV, is another example in the long history of magnet technology from early 20$^{th}$ century till now. The challenges here are not really on power efficiency or extreme uniformity but in generating a magnetic (B) field of about 1 Tesla (T) in the entire volume of the 50-100 ktons of steel. So the size or extent of steel over which this field is required is challenging from the point of view of fabrication and installation in the limited space of the cavern. As explained above, the steel in question is in the form of plates, one above the other, with active detectors interlaced between them. This leads to a large size and volume bounded by the ICAL magnet. Considering that the assembly of this magnet will proceed in the narrow confines of the cavern, the size of the steel plates that can be conveniently maneuvered and assembled to construct the magnet can be constrained. The other consideration is the ratio of the length of air gaps to the path length in iron for any layer. A smaller size of the tile results in a larger ratio making it inefficient while a smaller ratio due to a larger tile size results in difficulties in handling the plates. The choice of 4 m × 2 m × 0.056 m as the iron plate unit or tile to construct the iron layer was a good compromise between mechanical constraints imposed by cavern size and power economy in generation of magnetic field.

Since the air gaps between the iron plates, where the magnetic flux jumps across to the next plate is small in comparison to flux travel length in iron the power budget is expected to be small. This is in direct contrast to the conventional electromagnets which are used in the laboratory or accelerators where the B-field of interest is in air gap between the iron yokes. The iron core or yoke around the air gaps in these magnets provide a low reluctance path for the magnetic flux in the rest of the circuit [27]. In the ICAL magnet, although undesirable, we deal with small air gaps between steel plates which are tiled in certain way to get layer upon layer of magnetized steel in any module. Further, the action of cutting plates through thermal, water jet or optical means also adds a magnetically poor zone on to the air gaps that the magnetic flux has to face while jumping between adjacent plates. The optimal method of cutting and sizing the plates thus becomes very important as it has a direct bearing on the efficiency of ICAL magnet.

The design of the coils that run across the full height of any module is another unique feature that demands attention. Given the fact that the volume occupied by the coils is actually a dead zone for the particle detection, we need to optimize on the ampere–turn requirement on one hand and also find ways to construct such a large sized coil, and further, also find ways to assemble this with subsequent assembly of iron plates layer by layer to construct the full ICAL magnet. These mechanical aspects, while being very important, are not addressed in this paper.

## III. ICAL MAGNET DESIGN CONSIDERATIONS

The 51 kton ICAL detector consists of three modules each weighing ~17 kton. Fig. 1 shows the schematic of INO- ICAL

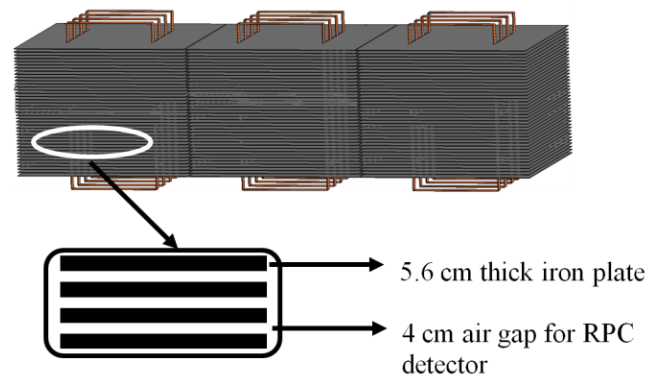

Fig. 1. Schematic of the INO ICAL detector with copper coils (in brown).

detector. The baseline ICAL magnet configuration for each of the three modules consists of 151 layers of low carbon steel. The thickness of each iron layer was decided on the basis of the following considerations. Given the cavern height a smaller plate thickness, for a fixed RPC (resistive plate chamber) detector thickness (see Fig.1), results in a smaller iron mass implying a lower neutrino event rate. On the other hand a larger plate thickness results in increased multiple scattering of the muon (resulting from neutrino interactions) and leading to a worsening in the momentum resolution. An optimum thickness was found to lie between 45 to 60 mm. We have chosen a nominal thickness of 56 mm for the iron plate. The resulting numbers of layers are adequate to have a reasonable efficiency for the muons in the energy range of ~1-20 GeV. The layers are alternated with gaps of 40 mm in which will be placed active detectors, of the RPC type, to measure the charged particles produced in ν interactions with the iron nuclei. RPC detector gives X, Y hit information at ~ 3 cm spatial resolution and timing information with a resolution ($\sigma_t$) 1 ns. Each iron layer has a dimension of 16 m × 16 m assembled from 2 m × 4 m tiles. The gap between consecutive tiles is ~2 mm to construct the full ICAL magnet that suits its assembly. Copper coils carrying DC current having a height (H) of ~15 m and width (W) ~8 m are used to magnetize the low carbon steel plates. The calorimeter will be magnetized with a uniform magnetic field (B = 1-1.5 T) to distinguish the $\mu^+$ and $\mu^-$ events from the opposite curvature of their tracks in the presence of a magnetic field.

### IV. SIMULATION OF MAGNETIC FIELD

The atmospheric neutrinos come from all directions, interact with nuclei of iron and produce muons and hadrons. The ICAL detector will be used to detect these muons and hadrons by reconstructing their energy and momentum. The magnetic field in ICAL will allow measurement of momentum and electric charge (from the sign of curvature of the track) of muons. The energy resolution also depends on the magnetic field and in general it improves with increasing magnetic field. Additionally magnetic field also increases the fiducial volume (FV) of the detector where FV is defined as an effective volume within a surface where one has 50% chance of contribution to events leading to a determination of the muon momentum. So the larger the FV of the ICAL magnet the higher is the useful fraction of events obtained. We have studied various configurations of the ICAL magnet and simulated the B-field in order to decide which is the most suitable for achieving the goals of ICAL.

#### A. B-field distribution study using various types of slots

The 3D electromagnetic simulation has been carried out for studying the B-field distribution and its uniformity by using Magnet software (Magnet 6.26 from Infolytica, Canada). The computation time to simulate the ICAL magnet having 151 layers of iron plate is very long. In order to reduce this time, a single layer of iron plate (16 m × 16 m × 0.056 m) with four and eight discrete and two continuous slots having 2, 2 (4) and 2 (4) coils, respectively, is used in the simulation. The coil dimensions are H ~ 0.456 m and W ~ 8 m. The cross section of the current carrying coil is 0.08 m × 0.625 m. We have checked that the simulated field with such a configuration agrees with that for a single central plane (within 4%) with that with coils of dimension 8m (W) × 15m (H) and also for 3 layers placed symmetrically at the top, middle and bottom. Figs. 2(a), (b) and (c) show the iron plate having discrete and continuous slots, respectively, through which copper coils pass to energize the ICAL magnet. The iron plate has eight discrete slots carrying two coils which are placed at the 1st and 4th pair of slots (slots pair are counted from bottom to top). The dimensions of the smaller sized slots, in Figs. 2(a) and (b) are 0.1 m (width) × 0.645 m (length) and the corresponding numbers for the continuous slots in Fig. 2(c) are 0.1 m × 8 m. The centre to centre longitudinal separation between two coils is 7.355 m for the iron plate with two discrete and continuous slots (each carrying two coils) and for the configuration with eight discrete slots carrying four coils is 2.452 m (same for continuous slots having four coils).

Slots are positioned at a distance (slot edge to iron edge 4 m in both X and Y-directions, respectively), shown in Fig. 2(a) which is (considered as baseline design) the reference slot configuration of the ICAL magnet such that magnetic lines of force should distribute symmetrically. The magnetic field (H) versus induction (B) data for two different soft magnetic materials M1 and M2 [28] (see Fig. 3) was used as an input for the electromagnetic simulation of the ICAL magnet. As can be seen material M2 is magnetically softer than M1 viz for a given B, M2 requires a smaller H-field than M1. The insert figure in Fig.3 shows the blown up part of the B-H curve at low H. The softness of the magnetic properties depends on the chemical composition with the carbon content being crucial.

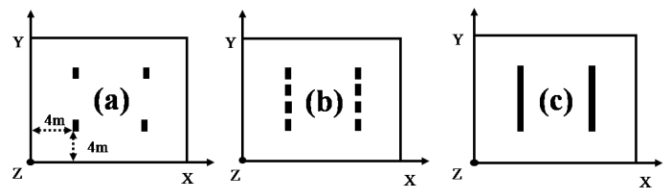

Fig. 2    The 16 m × 16 m iron plate having (a) four, (b) eight discrete and (c) two continuous slots (filled rectangular boxes)

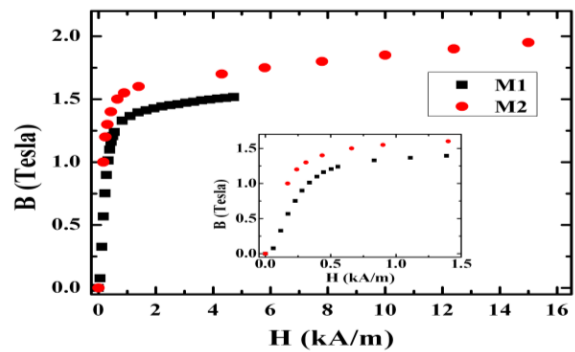

Fig.3.   Magnetic induction (B) versus field (H) for soft magnetic material M1 and M2 used in the simulation.

In order to address the estimation of the fiducial volume of the detector, the overall B-field distribution in iron (area of 16 m × 16 m) is needed. This is shown in Fig. 4 (a) for plates having continuous slots carrying four coils (5 kA per coil). The B-field has been averaged over a pixel size of 5 cm × 5 cm. The distribution is shown in Fig. 4(a). Fig. 4(b) shows the shaded plot of B-field and its direction and brings out the uniformity pattern of the B-field. The white circle and cross symbol show the outward and inward direction of current respectively. From the B-field distribution it can be seen that a lower field is found in the peripheral regions of the plate i.e. at four corners and at the middle (top and bottom) part of the plate, whereas, a higher field is found in regions which have proximity to the coils. From this, it is clear that all parts of the plate will not be equally effective for reconstruction of muon momenta and charge.

A comparative study on B-field distribution in iron due to various slots types are given in Table I. The iron plate has both four and eight discrete slots containing two coils give nearly the same fractional area for $|B| > 1$-$1.3$ T. For the case of an iron plate with eight discrete slots carrying four coils we see a larger fractional area of ~ 71 %, for $|B| > 1$ T compared to that containing two coils (~ 61 % for $|B| > 1$ T). This is due to the fact that when the iron plate has four coils, the high field region is distributed over a larger area compared to that of two coils. The reduced separation between consecutive coils also helps. An iron plate with continuous slots, on the other hand, gives B-field uniformity over a much larger area (~ 75 %) as compared to the plate with 4 (~ 61 %) and 8 (~ 61 %) discrete slots for $|B| > 1$ T with each of these configurations contains 2 coils. For the case of continuous slots, magnetic lines of force go around the slots (following a path of minimum reluctance) and get distributed more evenly in the plate compared to discrete slots.

TABLE I

COMPARTIVE STUDY OF B-FIELD DISTRIBUTION FOR ICAL MAGNET HAVING VARIOUS SLOTS AT 20 KA-TURNS

| No. of slots | Slots type | No. of coils | Current per coil (kA) | Fractional area (%) | | |
|---|---|---|---|---|---|---|
| | | | | $|B| > 1$ T | $|B| > 1.2$ T | $|B| > 1.3$ T |
| 4 | Discrete | 2 | 10 | 60.6 | 35.1 | 20.1 |
| 8 | Discrete | 2 | 10 | 61.1 | 35.1 | 20.1 |
| 8 | Discrete | 4 | 5 | 71.5 | 49.1 | 27.6 |
| 2 | Continuous | 2 | 10 | 78.7 | 40.9 | 17.5 |
| 2 | Continuous | 4 | 5 | 74.8 | 58.6 | 30.0 |
| 2 | Continuous | 1 | 20 | 70.8 | 54.9 | 26.8 |

It is also found that B-field distribution in the iron plate varies with coil configuration for plates having continuous slots carrying one (filled within 8 m of slot length, cross-section 0.08 m × 7.98 m), two (coil cross-section 0.08 m × 0.625 m) and four coils (coil cross-section 0.08 m × 0.625 m). It is observed that the plate with four coils gives a larger fractional area with $|B| > 1.2$ to $1.3$ T compared to the plate having one or two coils as shown in Table I. This is so because the centre to centre separation between two coils configuration (7.355 m) is more compared to that in the four coils configuration (2.452 m) which is less than half of the separation between two coils (for continuous slots having only two coils). In addition, the ICAL magnet has continuous slots where a whole slot filled with copper coils gives ~ 4 % less of the fractional area for which $|B| > 1$-$1.3$ T compared to the slot containing 4 coils. The coil field (H) which is the driving function for magnetization of the B-field in iron is more uniform with four discrete coils over that of a continuous coil [29].

From our study, we observe that slots having four sets of coils give a better B-field distribution in iron. In addition, increasing the number of coils decrease in current per coil as a result of which the ohmic losses reduce.

*B. Effect of slot length*

Since a large B-field will allow a better measurement of the muon momentum and charge, it is desirable to optimize the B-field value with respect to power. Studies have been carried out at various currents using configurations having continuous slots of length ($L_s$) 8 m as shown in Fig.5 (a) containing four coils in order to increase the B-field ($|B| > 1$ T) distribution in iron. It has been observed that, the fractional area for which $|B| > 1$-$1.3$ T increases with current beyond 20 kA-turns whereas at higher values ($\geq 1.2$ T) it increases with Ampere-turns though the increase is slow.

Our studies show that the configuration with continuous slots carrying four coils gives an acceptable B-field distribution over the 16 m × 16 m area. In order to try to improve this further, studies have been carried out for optimizing the length of the continuous slots, $L_s$, and their position (i.e. reducing the width ($W_s$) of side section) as shown in Fig.5 (a). The details of the comparative study are given in

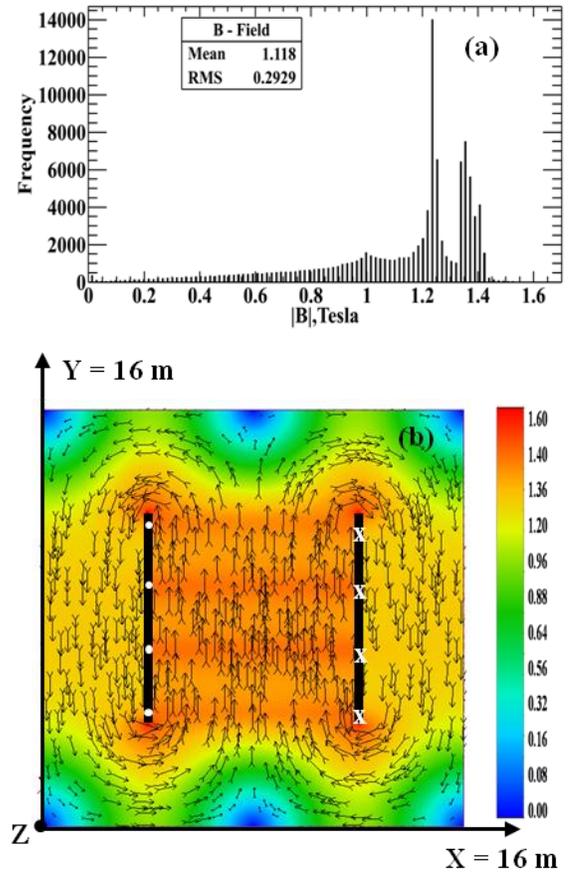

Fig.4 (a). Distribution of |B| -field in area of 16 m ×16 m, (b) shaded plot of B-field with arrows show its directions for magnet having continuous slots containing four coils at 20kA-turns.

Table II at two different Ampere-turns. It may observed that, at a given $W_s$ (3.95 m), B-field distribution in iron increases from nearly 66 % to 84 % with increase in $L_s$ from 7 m to 10 m for which $|B| > 1$ T at a current of 20 kA-turns. The increases of fractional area by increasing slots length is due to the fact that the lines of force turns around the slots and concentrate towards the edge of plate (minimum reluctance path). At 20 kA-turns, the fractional area for which $|B| > 1.2$ T increases from $L_s$ = 7 m to 9 m (similarly 7 m to 8 m at |B| > 1.3 T) and then decreases as path length followed by the lines of force around the slots increases and also the separation between coils increases. In addition, at $L_s$ = 7 m and $W_s$ = 4.50 m, fractional area reduces from ~ 46 % and ~ 23 % at |B| > 1 T

TABLE II

B-FIELD UNIFORMITY STUDY AT CURRENTS OF 20 kA-TURNS AND 60 kA-TURNS, 16 m × 16 m SINGLE PLATE FOR VARIOUS CONFIGURATIONS, d = CENTRE TO CENTRE SEPARATION BETWEEN TWO COILS

| $L_s$ (m) | $W_s$ (m) | d (m) | Fractional area (%), 20kA-turns | | | Fractional area (%), 60kA-turns | | |
|---|---|---|---|---|---|---|---|---|
| | | | |B| ≥ 1.0 T | |B| ≥ 1.2 T | |B| ≥ 1.3 T | |B| ≥ 1.0 T | |B| ≥ 1.2 T | |B| ≥ 1.3 T |
| 8 | 3.95 | 2.452 | 74.8 | 58.6 | 30.0 | 86.9 | 77.7 | 67.6 |
| 8 | 3.50 | 2.452 | 76.0 | 61.8 | 36.4 | 86.7 | 77.4 | 66.6 |
| 7 | 3.95 | 2.118 | 65.8 | 53.1 | 26.9 | 77.3 | 63.7 | 55.5 |
| 7 | 4.50 | 2.118 | 46.2 | 26.7 | 22.8 | 64.0 | 32.0 | 26.5 |
| 9 | 3.95 | 2.785 | 81.5 | 63.2 | 32.4 | 88.3 | 80.9 | 68.6 |
| 9 | 3.50 | 2.785 | 82.1 | 69.3 | 33.7 | 89.6 | 82.7 | 76.2 |
| 10 | 3.95 | 3.118 | 83.1 | 46.9 | 20.3 | 89.6 | 82.1 | 60.5 |
| 10 | 3.50 | 3.118 | 84.4 | 55.1 | 20.1 | 91.6 | 85.9 | 80.4 |

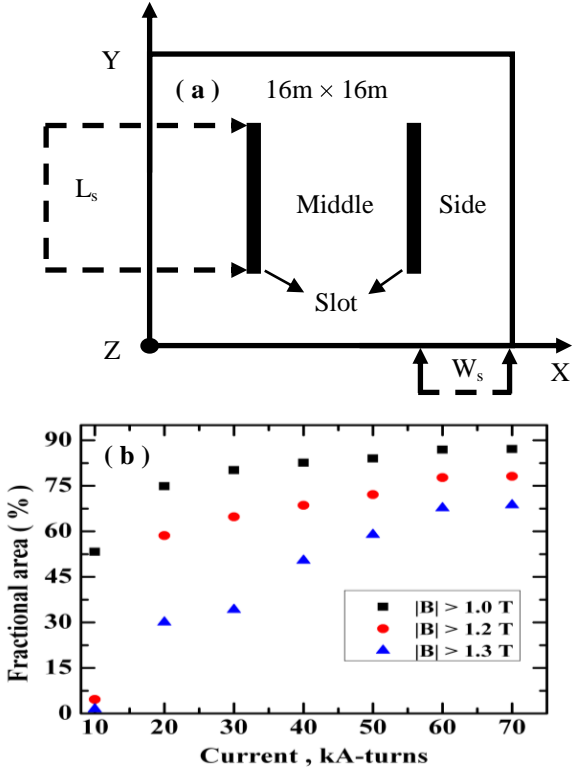

Fig. 5 (a). Schematic of ICAL magnet with different sections, $L_s$, $W_s$ is length of slots and width of side section respectively, (b) B-field uniformity studies at various currents for slot length of 8 m.

to 1.3 T respectively. This is due to lines of force following the minimum reluctance paths not evenly distributed in the plate. The fractional area for which |B| > 1 T to 1.3 T increases with decreasing $W_s$ for a given $L_s$. This is due to increase of B-field in a smaller area ($W_s$ × plate thickness). Similar behavior has been observed at Ampere-turns of 60 kA-turns. The configuration with slots having $L_s$ = 10 m and $W_s$ = 3.5 m seems to be the optimal with respect to the B-field distribution in 16 m × 16 m area of iron plate. The configuration with slots having $L_s$ = 9 m and width of side section $W_s$ = 3.95 m are suitable from a practical design standpoint and are close to the optimal value. From this study, it has been observed that by increasing the length of the slots and decreasing the width of the side section the B-field uniformity increases. In reality, the choices of the slots dimension and their position will also be constrained by the other mechanical properties of the ICAL magnet and the sizes of the detectors.

### C. Effect of plate thickness on B-field distribution

At low energy, the momentum resolution of muons is affected by multiple Coulomb scattering which increases with increase in plate thickness. At high energies, this resolution becomes poorer and there is decrease in charge identification efficiency due to the reduction in bending of particle trajectory.

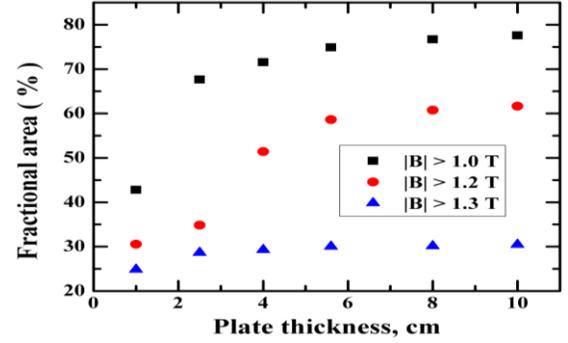

Fig. 6. B-field uniformity study at NI = 20 kA-turns, 16 m × 16 m × 0.056 m single plate for various plate thickness.

This study has, therefore, been carried out using the "baseline" ICAL configuration (continuous slots with four coils) where only the plate thickness has been varied. The current per coil is 5 kA. It has been found that the fractional area for which |B| > 1 T increases with increase in plate thickness as shown in Fig. 6 and saturates for plates having thickness beyond about 4 cm. This is because the magnetic lines of force get redistributed in the iron due to the increase in plate thickness. It has also been found that fractional area for which |B| > 1.3 T shows saturation behavior with increase of plate thickness. Similar behavior is found for total current of 40 kA-turns (10 kA per coil). Finally, the dependence of field uniformity with plate thickness reinforces our choice of 56 mm as the design value for plate thickness.

### D. Effect of soft magnetic properties of material on B-field distribution

A comparative study has been carried out for the B-field distribution using two different soft magnetic materials as shown in Fig.3. At a given magnetic field (H), material M2 is softer than M1. In the electromagnetic simulation of the ICAL magnet the corresponding B-H data is used as input in the ICAL magnet simulation. The comparative study has been carried out by considering single layer of iron plate (16 m × 16 m × 0.056 m) with continuous slots carrying four coils. It has been found that, an ICAL magnet made up of material M2 gives a larger fractional area for which |B| > 1 T compared to that with M1 for the same Ampere-turns, as shown in Fig. 7. In particular at lower Ampere-turns (10 kA-turns), material M2 gives more fractional area (~ 79 %) compared to M1 (~ 53 %) and at higher current (60 kA-turns) the difference between them reduces (material M2 gives ~ 90 % and M1 ~ 87 %) for which |B| > 1 T. Both materials show the saturation of the B-field at higher currents. From this study, it can be seen that, the material property decides the Ampere-turns required for getting the higher B-field distributions. In practice, though, the availability and cost might be the overriding factors.

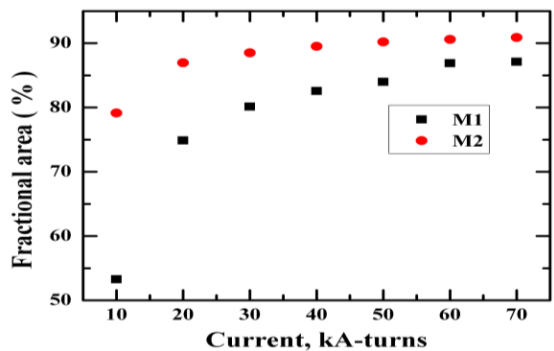

Fig. 7. Comparison of variation of fractional area with |B| > 1 T with current using two different soft magnetic materials M1 and M2 (single layer magnet 16 m × 16 m × 0.056 m).

### E. Effect of tiling and gaps on B-field distribution

All the studies reported in this paper have been carried out by considering single layer of 16 m × 16 m × 0.056 m iron plate. Practically, however, building the ICAL detector using this size plate is not feasible due to difficulties in manufacturing and handling. So the 16 m × 16 m area has been tiled with plates of size 2 m × 4 m. Due to mechanical tolerances, there will be air gaps between the tiles. The magnetic lines of forces fringe out at air gaps which lead to the reduction of flux linkage amongst them. As a result more Ampere-turns are needed for getting the required B-field. Instead, if the size of the tile is reduced, the number of gaps between them will increase leading to an increase of the fringing flux in the gap, and hence, in the exciting current for a required average B-field. An increased tile size leads to difficulties in handling and edge machining. For simulation, a gap of 2 mm among the tiles has been assumed as shown in Fig.8 (extended view) and tiles are arranged in various ways to get the maximum B-field at minimum Ampere-turns.

The magnetic field in iron for various current has been studied using four different configurations C-1 to C-4. Fig. 9 shows that configurations C-1 and C-4 give the same $B_y$ (point of observation is nearly middle of the plate where B-field is almost in the Y-direction) for all Ampere-turns and is similar for C-2 and C-3. The configuration C-2 gives ~14 % and ~ 2 % more $B_y$ at total current of 20 and 50 kA-turns respectively, as compared to C-1. This is due to a smaller number of air gaps seen by magnetic lines of force in a closed loop which results in less fringing (leakage) of flux in which C-2 configuration appears to be optimum. In this configuration, the effect of air gap on the |B|-field in iron is shown in Fig.10 for gaps of 0 mm (single plate, no tiling structure), 2 mm, 4 mm and 6 mm. The flux density in iron decreases as the fringing flux increases with increase of air gaps. At lower current the reduction in B-field is higher than that at higher values of current.

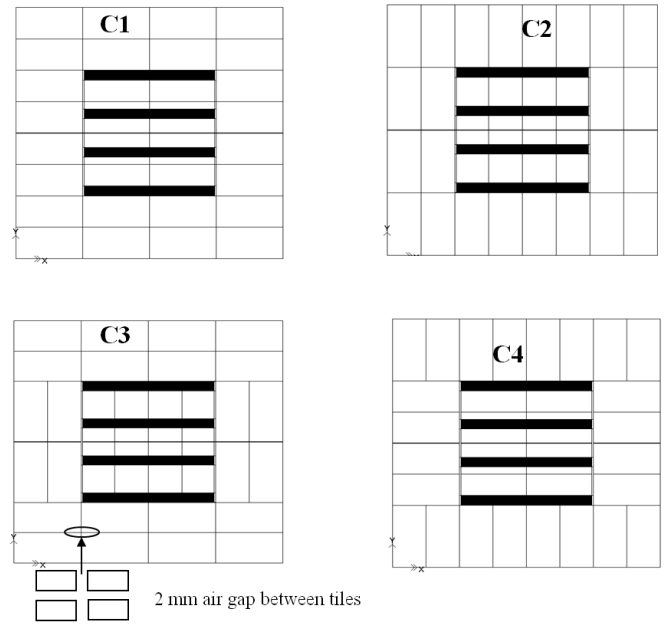

Fig. 8. ICAL magnets with various arrangements of tiles, dark black color shows the coils (Top view).

### F. Comparative study of B-field distribution using C-2 configuration and single layer of iron plate

It has already been seen that the B-field in iron reduces due to increase of air gaps among tiles. A comparative study has been carried out to find out B-field distribution using a single layer (no tiling structure) of iron plate (16 m × 16 m × 0.056 m) and layer with tiling configuration (C-2 configuration, gap between tiles is 2 mm, 4mm, 6 mm).

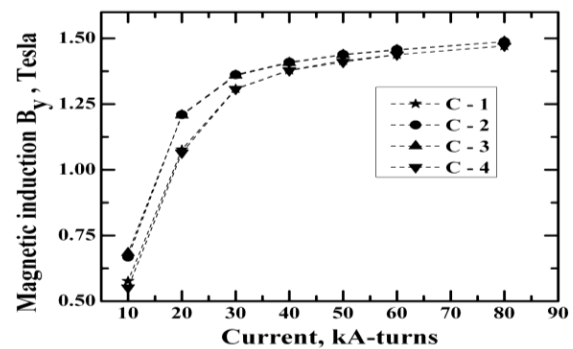

Fig. 9. B-field variations with currents for different configurations.

It can be seen that fractional area for which |B| ≥ 1 T reduces by increasing of gap between tiles with respect to a single plate with Ampere-turns as shown in Fig.11. This is due to fringing of flux occurs at gaps among the tiles. The reduction in field distribution is more at low current (10 kA) i.e. single plate gives ~ 53 % where for tiling structure of various gaps the fractional area for which |B| > 1 T is very small and the difference between single plate and tiled structure reduces for higher current. This shows that the smaller the gap between the tiles the large fraction of area of ICAL magnet can be used for studying the muons. Taking into account the mechanical constraints, we have chosen a gap of 2 mm for ICAL.

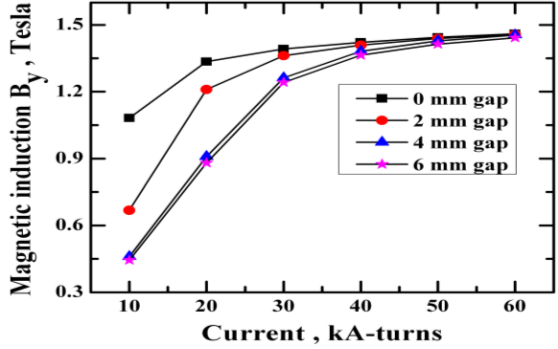

Fig. 10. B-field variation with current at different gaps between tiles for C-2 configuration.

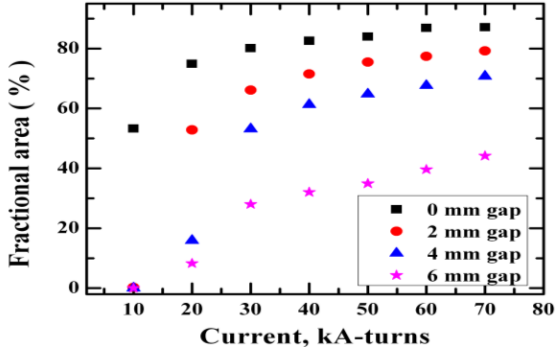

Fig. 11 Comparison of variation of fractional area with |B| > 1 T with current for single layer of ICAL magnet and C-2 configuration. The gap between tiles is 2 mm.

## V. MAGNETIC FIELD RESPONSE OF MUONS

One of the principal goals of the INO ICAL detector is to address the neutrino mass hierarchy. For this it is important that the neutrino energy and incoming direction be accurately measured in each event. The energy of neutrino is sum of muon and hadron energy for charged current events. The muon energy can be measured either from curvature in a magnetic field or from the range of muons stopping in the detector or both. It is difficult to reconstruct the energy of individual hadrons. However, the hit multiplicity of charged particles distinct from the muon track can be used to calculate the total energy of hadrons in each event.

We have studied the effect of the strength of the B-field on the energy resolution and charge identification (ID) efficiency of muons. An object oriented "detector description and simulation tool", GEANT4 [26], is used for the detailed simulation of the ICAL detector response to muons and hadrons [25] resulting from ν interactions with iron nuclei. The magnetic field map in a grid size of 5 cm × 5 cm of the iron plate obtained from the above electromagnetic simulation of ICAL magnet has been interpolated for the 16 m × 16 m and used for muon tracking. A sample of $10^4$ muons ($\mu^-$) with energies 1- 20 GeV originate in the central region of the detector where the B-field is nearly uniform as shown in Fig. 4(b), in all azimuthal directions (Φ = 0 to 2π) and at zenith angles ($\theta_z$) such that cos $\theta_z$ takes on values between 0.1 to 1.0 in bins of Δcos $\theta_z$ = 0.1. An algorithm based on the Kalman Filter [30] is used to reconstruct the muon momentum by considering the tracks close to the vertex (with a cut on $\chi^2$/number of degrees of freedom < 10). The reconstructed momentum distribution obtained from the simulation is plotted in the range of 0 to 2 $P_\mu$, where $P_\mu$ is the input momentum (GeV/c). Fig. 12 shows the distribution of the reconstructed momentum ($P_{rec}$) for 5 GeV muons incident at a zenith angle corresponding to cos $\theta_z$ = 0.65 and for all azimuthal directions. This reconstructed momentum distribution is fitted to a Gaussian function in the range of $P_\mu$ - FWHM (full width half maximum) to $P_\mu$ + FWHM. The dashed line shows the fitted Gaussian function. The standard deviation (σ) obtained from the fit is used to extract the energy resolution of muons. The charge identification efficiency of muons for ICAL detector has been estimated within the range of 3σ. From the figure it can be seen that the reconstructed mean value agrees quite well with the input value.

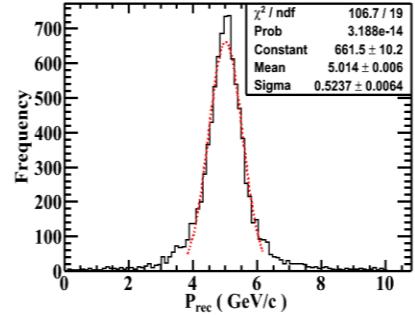

Fig.12. Reconstructed momentum distribution of 5 GeV muons at cos $\theta_z$ = 0.65. The dashed line shows the Gaussian fit to the distribution.

The energy resolution of muons is affected by multiple scattering, bending (sagitta) of the trajectory and the number of hit points used to fit the track [31]. At low energy, the resolution is mostly affected by multiple scattering and this reduces with increase of energy. Fig. 13(a) shows the variation of muons energy resolution with energy at magnetic fields of 1.1 T, 1.5 T and 1.8 T (where this refers to the central region of the detector) are incident at cos $\theta_z$ = 0.65. To get magnetic field of 1.8 T we have used material M2 as shown in Fig.3. For a given magnetic field, the energy resolution improves with increase in energy. The best resolution is at around 5

GeV and becomes poorer thereafter. At low energy ($E_\mu \leq 5$ GeV), the energy resolution improves with increasing energy as the number of hit points increases. At higher energy, the poorer resolution can be attributed to more uncertainties in the sagitta. Apart from this, the poorer resolution also arises due to reasons such as muons passing through the spacer of the detector or the trajectory falling in the gap between two RPC detectors. For a given energy, the resolution improves with increase in magnetic field due to increase in bending of trajectory resulting in a reduced uncertainty in the measurement of the sagitta. It has been observed that, at the higher energies, the energy resolution improves by ~ 21% to ~ 29% for energies about 9 GeV to 20 GeV, when the magnetic field is increased from 1.1 T to 1.8 T for muons incident at a zenith angle corresponding to cos $\theta_z$ = 0.65. The higher magnetic field can be obtained by increasing the current in the coils or using softer magnetic materials. These will increase the running cost of the detector.

function $\frac{a}{E^\alpha} + bE$, where "a" arises from multiple Coulomb scattering which is important at very low energies and "b" is due to the error in position measurement which is important at high energies, where the bending is very small (leading to a larger error on sagitta measurement). The parameter "α" takes a value of 0.5 to 0.6. Nevertheless it can be seen that the resolution does not improve significantly in the energy range of 1 to 20 GeV by increasing the B-field from 1.5 to 1.8 T. Since INO physics is most sensitive to the neutrino in the few GeV range, therefore, choice of 1.5 T B-field seems quite appropriate.

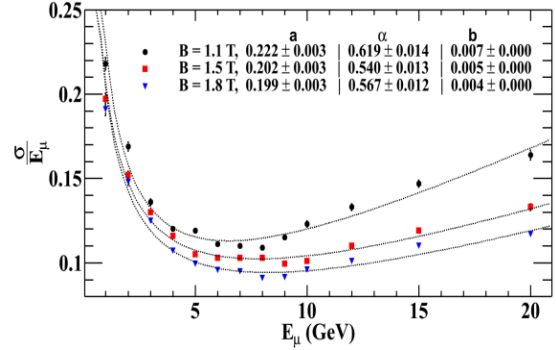

Fig.14. Energy resolution at **B** = 1.1 T, 1.5 T and 1.8 T is fitted to a function $\frac{a}{E^\alpha} + bE$ (dashed line).

## VI. SUMMARY

The electromagnetic simulations of the ICAL magnet have been carried out to find the optimum field values and design parameters. The configuration having continuous slots for accommodating the coils gives a superior magnetic field uniformity compared to those with 2 or 4 pairs of slots. The B-field distribution in iron increases with increase of plate thickness and gets saturated beyond the plate thickness ~ 4 cm justifying the choice of 5.6 cm. It has also been found that the B-field distribution depends on the magnetic properties of the material. In particular near saturation fields can be achieved using softer material at much lower power consumption. However it remains to be seen whether large scale manufacture and procurement of this steel is feasible, which is being explored now.

The effect of the four tiling arrangements on the field distribution and power consumption has also been studied. The particular configurations C-2 and C-3 appears to also give better results compared to C-1 and C-4. The effect of size of the air gap between adjacent tiles leading to a reduction of the B-field for a given excitation currents has also been seen and where the smallest gap is desirable, a gap of 2 mm may be practical.

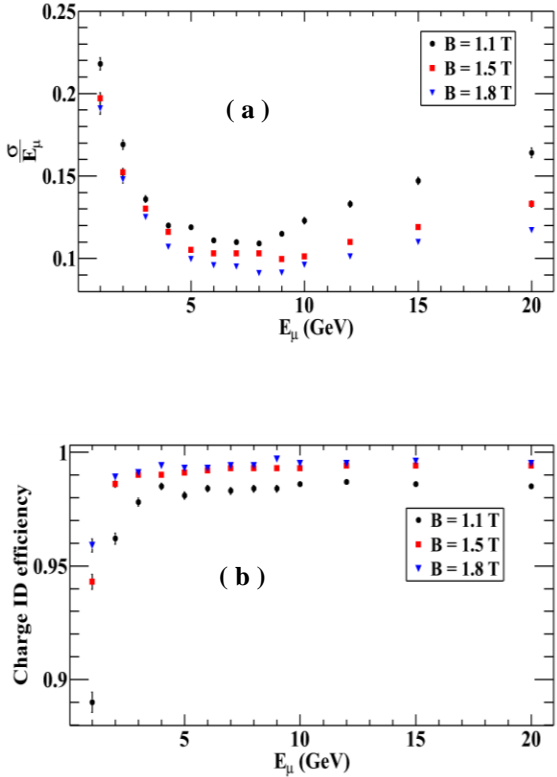

Fig. 13 (a). Resolution and (b) charge identification efficiency versus energy of muons at cos $\theta_z$ = 0.65.

Fig.13 (b) shows the variation of charge identification efficiency of muons with energy at different magnetic fields. It shows that the charge identification efficiency is more than 90 % for all energies and increases with increasing magnetic field. A similar behavior for energy resolution and charge identification efficiency has been observed for a range of values of cos $\theta_z$ between 0.95 and 0.15.

In Fig. 14, the muon energy resolution (cos $\theta_z$ = 0.65 and averaged over all the azimuthal angles) has been fitted with a

At a given magnetic field, the energy resolution improves with increasing muons energy. At lower energies the resolution is affected by multiple scattering while at higher energies the poorer resolution is attributed to the uncertainty in sagitta measurements. The lowest energy resolution obtained is ~ 10 % for 5 GeV muons at a magnetic field of 1.5 T. It has been observed that the energy resolution as well as charge identification efficiency of muons improve with the increase of the magnetic field, but a choice of 1.5 T B-field appears to give satisfactory results and not much gain was attained by increasing the field to 1.8 T.

## ACKNOWLEDGEMENTS

We would like to thank P. Verma, S. V. L. S. Rao, Suresh Kumar and Y. P. Viyogi for their helpful suggestions and useful discussions. We would also like to thank Gobinda Majumder and Asmita Redij for developing the INO-ICAL GEANT4 detector simulation and reconstruction packages. We are thankful to Md. Nayeem, M.V.N. Murthy and D. Indumathi for a critical reading of the manuscript. S. P. Behera would like to thank Meghna K. K. for discussions on GEANT4.